# Cyber Attack Surface Analysis of Advanced Metering Infrastructure

James Christopher Foreman
School of Engineering Technology
Purdue University
West Lafayette, IN, USA
jchrisf@purdue.edu

Dheeraj Gurugubelli
Dept. of Computer and Information Technology
Purdue University
West Lafayette, IN, USA
dgurugub@purdue.edu

*Abstract*— Cyber attacks on critical infrastructure have been an issue of importance in industry over the past several years, as well as a focus of academic research. Cyber attacks of various types and magnitude have been on the rise targeted specifically at power grids. A successful attack on the power grid could result in significant impacts including grid shutdown, cascading failures, damage to the infrastructure, and potential harm to people. Power grid infrastructure are critical in nature; they enable operations for residential, commercial, industrial, and government users across critical infrastructure such as water, communication, banking, transportation, manufacturing, and more. The compromise of these operations introduces threats, which span from economic to public safety. The Advanced Metering Infrastructure (AMI) is now being rapidly deployed throughout the power grid, and is an enabling technology for smart grid. Identifying the attack surface is a necessary step in achieving cyber security in smart grids and AMI. The purpose of this paper is to quantify and examine the cyber attack surface of AMI in order to lay the foundation for mitigating approaches to AMI cyber security.

I. INTRODUCTION

The security of Advanced Metering Infrastructure (AMI) has recently become an area of interest as its deployment has grown. The affinity towards AMI by utility companies, energy markets, and regulators is primarily to facilitate near real-time collection of power flow and usage data. This will allow utilities to provide dynamic pricing services, demand response, and perform better management of the power grid, although these new abilities increase the attack surface [1,2]. Smart meters are installed at homes, buildings, and other facilities of the power consumer and thus, the vast number in deployment will be in the many of millions [3,4].

Like any nascent system, AMI have yet to establish security measures to handle cyber attacks beyond rudimentary measures commonly employed in general, e.g., network encryption. A cyber attack is an attempt by hackers to damage or destroy a computer network or system. Attacks can vary in complexity, magnitude, and impact. A cyber attack on AMI may involve intelligence gathering, infecting the target AMI systems, AMI exploitation, exfiltration of data from various attack points of AMI, and maintaining control [5,6,7]. The cyber attack surface can be defined by the methods an environment or a system can be attacked by an adversary to introduce or retrieve data from that environment or system. A targeted attack on AMI could potentially result in shutdown of the power grid, disabling energy delivery systems [8]. This could have devastating effects on government, trade, commerce, banking, transportation and other important aspects, which rely on energy to operate [2]. A compromise of AMI may also result in an invasion of privacy [9], and provide a surface from which to extract information from users such as Internet activity, financial, or health records.

Cyber security of AMI was recently a topic of major interest at a Black Hat conference where some vulnerabilities and simulated attacks were demonstrated [10]. Yi et al [11] demonstrated a specific denial of service attack where an attacker may select any node and disrupt traffic on the AMI network. Attackers may also compromise a smart meter to change energy usage data or fabricate other meter data [12]. Cleveland [13] discussed an example of a possible scenario of a smart hacker cracking AMI security and sending 5 million remote disconnect commands. Attacks on critical infrastructure in general have been on the rise and a survey [14] conducted by McAfee indicates in one year's time that one in four have been the victims of cyber extortion or threatened cyber extortion; denial of service attacks had increased from 50% to 80% of respondents; and approximately two-thirds have found malware designed to sabotage their systems.

The key contribution of this paper is the timely study of the AMI attack surface given the rapid deployment of AMI. As of 2012, over 43 million smart meters have been installed in the U.S., 89% of which are residential with the remainder to commercial and industrial consumers [4]. According to the Institute for Electric Efficiency smart meter deployment projections, approximately 65 million smart meters will be deployed in the U.S. by 2015 [3]. The cost of AMI is huge given this scale, and frequent replacement with more secure units is extremely cost prohibitive. This research adds value in a concurrent way by aiding understanding of AMI cyber security exposure, which enables utilities to be more informed regarding the security posture of AMI.

## II. Previous Work in AMI Attack Surface

Security concerns with respect to the power grid resulted in a set of compliance standards developed by the North American Electric Reliability Corporation (NERC). These are the Critical Infrastructure Protection (CIP) standards, which was a first step towards the cyber security compliance regulations for power grids [15]. These standards primarily focus on the identification and protection of cyber critical assets in the power grid infrastructure. NIST has also developed a set of standards for the interoperability of assets comprising the smart power grid, i.e., *Framework and Roadmap for Smart Grid Interoperability Standards* [16], which briefly addresses the need for cyber security standards and the components of these standards. An overall strategy for implementing cyber security in AMI is also presented in Chapter 6 of this text, but only from a generalist point of view and with no analysis of the attack surface or specific vulnerabilities.

Schukat [17] attempts to achieve a holistic view of the cyber threat landscape in smart grid infrastructure. His research focused on Machine-to-Machine (M2M) communication, the role of PKI, and authentication protocols used for mitigating cyber risk. He discusses that energy systems are going through a paradigm shift towards better and smarter solutions driven by factors like economic cost, efficiency, environmental impact, etc. The smarter the systems get, the more complex they got. He argues that the IP-based systems and open standards in the modern energy systems, whilst beneficial for many reasons, pose potential risk to the attack surface of energy systems.

Hahn and Govindarasu [18] proposed a security model, which is capable of representing various privilege states and discusses the various attack paths an attacker could take. Using this model, they attempted to produce an exposure metric that could help evaluate the completeness of security mechanisms in place in the smart grid. They discussed the necessity to support the grid and its security mechanisms in place through developing a security model, which would concentrate on the critical information necessary to protect the grid. They also mentioned that the complexity of the AMI infrastructure and constrained network bandwidth could limit the security monitoring. Their proposed exposure analysis algorithm could be used to quantify the exposure of the information objects within the smart grid. Though their paper could be helpful in providing quantitative metrics to evaluate risk in smart grids, it does not thoroughly discuss the complete surface attack of AMI.

Searle and McParland [19] at Lawrence Berkeley National Laboratory attempted to provide a non-proprietary insight into the AMI systems and smart meters within AMI. He further discusses some security issues and vulnerabilities that could lead to potential attacks. Accordingly, studying both the components of AMI systems and the higher architecture will promote meaningful and open understanding of the cyber security risks involved by installing smart meters in HAN [19]. Though this paper briefly discusses the architecture and components of AMI, its primary focus is limited to potential cyber attacks at the HAN level, a thorough discussion of the Smart Energy Profile (SEP), and security vulnerabilities in the communication standards.

## III. AMI as Critical Infrastructure

The AMI is the point of interface between the utility power provider and the end consumer of power. As such, it is a critical point in the power delivery system. The AMI can provide valuable information to the utility including real-time voltages, currents, power flow, power factor, etc., as well as tracking accumulated power usage by the consumer for billing purposes. The AMI also provides power control by connecting and disconnecting the consumer from the power grid. Attacks that compromise the integrity of the AMI can result in a loss of situational awareness and even control of the power grid by utilities [2]. Such attacks also have the potential to cut power from consumers, which include homeowners, businesses, hospitals, and other critical infrastructure such as water and telecommunica-tions [2]. The deployment scale of AMI results in many thousands to millions of users within a power utility's network, so the potential for widespread disruption and harm is significant.

Illustrated in Fig. 1 is an example of a power grid network topology. The AMI is intimately integrated with the power delivery process and the power grid, which is comprised of physical devices, such as transformers and circuit breakers. Disruption in AMI can create instability in this physical process potentially damaging equipment and causing physical harm to individuals utilizing the power grid. Smart meters comprising the AMI are distributed throughout a metropolitan area, usually serving thousands to millions of users. The effects of attacks can even migrate to other utilities and metropolitan areas, perhaps even on a national scale.

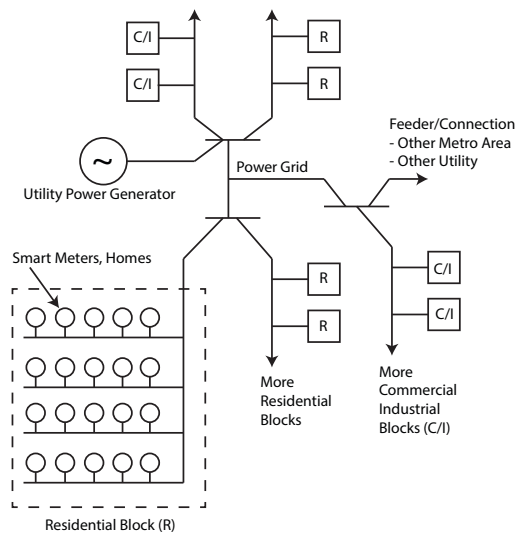

Figure 1 – Power grid network topology.

Types of attacks that are directed at the power grid itself include the following:

- *Denial of Power* (DoP) – Attacks on individual smart meters to deny power service to a consumer. This type of attack could be directed at critical users resulting in significant harm.
- *Theft of Power* (ToP) – Attacks on individual smart meters to facilitate theft of power from the utility. Consumers that are purposefully disconnected from the power utility may surreptitiously re-connect to obtain power. The data within a smart meter may also be altered to misrepresent power usage on any number of smart meters.
- *Disruption of Grid* (DoG) – If a large group of smart meters are compromised, they can be connected and disconnected in rapid or chaotic sequences, which results in instability in the power grid. On a large enough scale, the power grid will be unable to absorb this transient behavior and may partially fail resulting in widespread power loss.

## IV. AMI ATTACK SURFACE

### A. Smart Meters

Smart meters are the primary point of data collection for power grid energy consumption. They facilitate a dense and large-scale metrology of the power grid operating characteristics of voltage, frequency, power factor, etc. Lastly, smart meters also facilitate the automated connection and disconnection of consumers to the power grid. Figure 2 illustrates the block diagram of a typical smart meter with potential attack vectors.

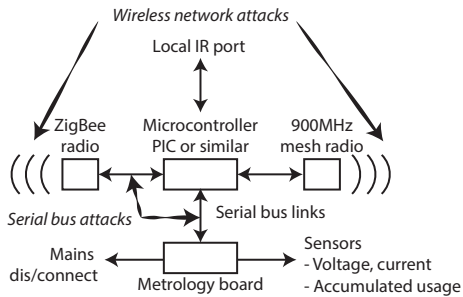

Figure 2 – Typical smart meter block diagram.

The design of the smart meter incorporates two network communication radios, and often a local infrared serial port for maintenance. This results in a broadened attack surface. Physical access to the meter adds another attack surface since internal serial links are typically unsecured and based on common protocols. Essentially, every smart meter becomes a multi-network terminal on the AMI information network. The microcontroller, radios, and firmware are driven by low cost, and as such minimal capability, since most utilities will need to purchase smart meters by the millions. Smart meters send usage statistics to utility companies at a relatively high refresh rate of minutes to seconds. Coupled with their vast number in deployment, an immense flood of big data is made available to the utilities. This enables dashboarding of power grid operation, automated operation and recovery, dynamic pricing, and more usage-based services to the consumer, but at the cost of privacy and vulnerability to cyber physical attacks.

### B. AMI Information / Communication Network

The AMI communications network is primarily comprised of the smart meter, the Smart Meter Data Collector (SMDC), and the network linking these, typically a wireless Frequency Hopping Spread Spectrum (FHSS) mesh or other cellular network. The AMI information network also includes the link to the Home Area Network (HAN) on the consumer side, typically WiFi or ZigBee®, and the link to utility's WAN, typically Ethernet infrastructure. The AMI communications network is geographically distributed throughout a metropolitan area alongside the power grid. The scale of this network can consist of hundreds to thousands of SMDC, each capable of serving thousands of smart meters, thus many thousands to a few millions. This is illustrated in Fig. 3 with potential attack vectors.

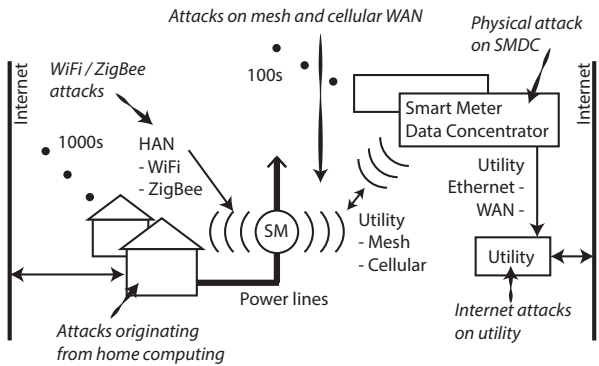

Figure 3 – AMI information network.

The vast number of devices coupled with the simplicity of individual smart meters results in a huge network of highly similar actors that are typically ill equipped to handle cyber security technologies for attack prevention. Another complication given the cost of deploying such a large number of devices is that devices are expected to have a long lifetime, perhaps twenty years or more, which means that hardware will remain constant for long periods of time. This contributes to a long lifetime for vulnerabilities even once known since upgrades are difficult or limited in scope. Therefore, the compromise of even a single smart meter through focused attack or reverse engineering potentially provides access to the AMI network as a whole. It is also interesting to note in Fig. 3 that the Internet bounds the AMI information network on both sides. This, coupled with the extensive use of multiple wireless technologies and geographic dispersion, results in an attack surface of unprecedented scale.

## C. Home Area Network

The HAN is the consumer side of AMI. The HAN begins with a wireless link, typically ZigBee, from the smart meter to a consumer gateway device in the home. The consumer gateway also acts as a bridge between the smart meter ZigBee network and the consumer's home devices, such as smart appliances and energy monitors. The consumer gateway will incorporate the ZigBee and Wi-Fi radios in addition to a Linux microkernel computer, usually containing an embedded web server. The gateway collects energy usage data, network status and text messages from the utility for display to the consumer on the gateway itself or a remote monitor. The gateway also forwards demand-response and energy-pricing signals to the smart appliances for their information. This gateway may also serve a connection to smart appliances, whose operating parameters can be influenced by the smart meter. For example, smart clothes dryer can be configured to dry clothes only when the price or demand for energy is low.

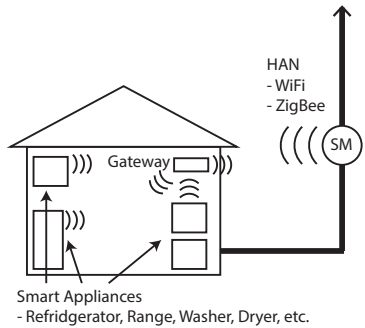

Figure 4 – Consumer gateway in AMI network.

The motivation for consumers to participate is additional energy cost savings in an intelligent, automated, and real-time approach. The consequence of this benefit is a new attack surface through these embedded devices, which typically have reduced processing power to incorporate sufficient cyber attack mitigating technologies.

The HAN is also the network that interfaces with the consumer's home computing systems and Internet connection. Consumers will likely want to monitor and collect information, perhaps even enabling Internet access by the utility for advanced energy usage statistics on individual appliances, or by themselves to remotely control appliances when away from home. A compromise of the HAN side of AMI has the potential to disrupt the lives of consumers on an intimate level, physically within their home, and provides an access point back to the Internet. This model also extends to commercial and industrial users who want the same capabilities for their enterprise. Fig. 5 demonstrates some of the probable attacks through the HAN side of AMI.

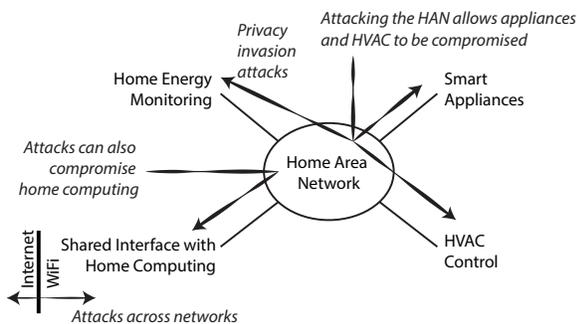

Figure 5 – HAN devices and probable attacks.

## D. Smart Meter Data Collectors and Head End

The Smart Meter Data Collector (SMDC) is a hardware computing device, based on a Linux microkernel, that aggregate the massive real-time data from a bank of smart meters by providing a collection and management point for the utility. Typically, a few thousand smart meters are managed per SMDC utilizing multiple FHSS mesh radios. The SMDC then provides a link directly to the utility via hardwired Ethernet or some other WAN backbone. The SMDC is usually responsible for cryptographic key management for the smart meters as well. Their position in the AMI information network is illustrated in Fig. 3 and the block diagram for a typical SMDC is shown in Fig. 6. The SMDC is subject to the same types of attacks as the smart meter.

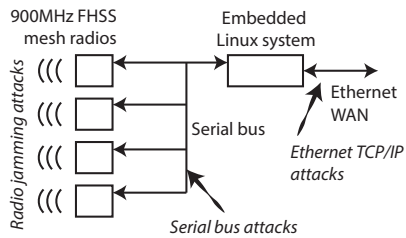

Figure 6 – Smart meter data collector.

The AMI Head-End (AMI-HE) is the AMI management platform at the utility installation. The AMI-HE centralizes management of the whole AMI and provides data warehousing for all collected data. This is also the interface point for the utility infrastructure from which billing, Outage Management Systems (OMS) for restoration, Energy and Distribution Management Systems (EMS/DMS) for dispatch and system monitoring, and Meter Data Management Systems (MDMS) from which other AMI management functions derive. The AMI-HE is typically wholly contained within the corporate WAN as an enterprise computing platform and as such, is secured by typical cyber security policies practiced in corporate systems. As such, the focus of this paper will remain within the AMI information network depicted in Fig. 3.

*E. Protocols and Software*

As illustrated in Fig. 3, AMI utilizes several different communication links and subsequently, several protocols, all of which provide an attack surface. AMI is enabled through the ANSI C12.22 standards, which define data messaging standards for smart meter AMI networks. The protocol describes transporting C12.19 table data [20] for interoperability among AMI devices with AES encryption. The attack surface of this protocol includes weak encryption, key discovery through reverse engineering or physical access, as in Figs. 2 and 6, which affords access to the data tables allowing reconfiguration, data collection/corruption and other control functions.

The ZigBee Alliance's Smart Energy Profile 2.0 Public Application Protocol Specification (SEP 2.0) [21] is a set of interoperability standards to connect consumer-side HAN devices to the smart grid. Illustrated in Fig. 7, is an illustration of the SEP 2.0 stack and potential attacks on its attack surface.

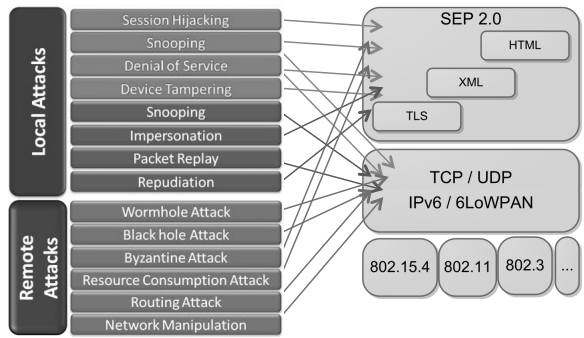

Figure 7 – SEP attack surface, *adapted from* [22,23].

Many software tools for hacking and cracking AMI are becoming prevalent such as the KillerBee framework and toolset [24]. The framework allows users to build their own tools for eavesdropping, emulating devices, etc. for exploiting ZigBee and IEEE 802.15.4 networks.

Grochocki et al [25] investigated attacks on the smart meter to data collector (SMDC) mesh radio network, including the development of an attack tree. Some of these attacks have been illustrated in Fig. 6 on the SMDC itself. The attack surface of the RF communication link is of particular interest since the SMDC interfaces with many thousands of smart meters and access to the device can be easily obtained physically through Ethernet and unsecured serial as well as medium range RF where radios are readily available to facilitate such attacks, e.g., WiMAX, 3G/4G LTE, cellular, etc.

It should be noted that many AMI devices rely on secret communication codes and algorithms embedded in the hardware with the hope of maintaining security through obscurity [25]. There has been some difficulty in cooperation between vendors and cyber-security researchers in an effort to protect intellectual property and competitiveness.

However, physical and/or RF access to the smart meters and SMDCs can expose these codes with some knowledge of hardware and reverse engineering [11]. Once exposed, the obscurity aspect is negated, along with any security afforded. According to a researcher at IOActive, all it takes for an attacker to gain command and control over a smart meter is $500 worth of equipment and materials and a background in electronics and software engineering [11]. This could mean en masse manipulation of service to residential and business facilities.

## V. Motivation of Attackers

Motivation is an important factor in order to understand and respond to cyber attacks. Cyber attacks targeted towards AMI are usually motivated by terrorism to disrupt a nation's critical infrastructure. Even casual crackers/hackers seeking to test their skills are attracted to such a prominent target. Industrial espionage and other business disruptions, whether politically motivated or not, are another key motivation. Additionally, some attacks may seek to utilize AMI as a launching point for other Internet attacks by virtue of the vast number of devices, their ability to initiate many-pronged attacks, and even their ability to hide malicious or criminal data through dispersion among many nodes [26]. According to research by Ponemon, the organizations managing the U.S. critical infrastructure facilities are not well prepared for the attacks, if any take place [27]. All nations are vulnerable without making the smart meters resilient. Smart meters have been game changers for cyber attackers, affording them a rich new resource for widespread disruption [28].

Beyond the DoP, ToP, and DoG attacks on the power grid introduced in Section III, symptoms of AMI network attacks include: unauthorized web pages posted on Internet-connected web servers collecting data, outbound data transmissions using unknown protocols and ports, huge compressed file transmissions over AMI networks, unusual data load between data collectors and the smart meters, and unusual log entries.

## VI. Conclusion

AMI is an enabling technology for new initiatives in electrical power and its deployment plays an intimate role in electrical power delivery. Given the recent paradigm shift towards rapid and large-scale deployment, the attack surface is rapidly increasing. AMI equipment vendors have also taken an approach largely based on closed and proprietary security solutions in a false effort to maintain security, which deters evaluation and further research by the scientific community – although there are signs that this trend is slowing changing towards open standards [29]. A cyber security roadmap for AMI systems can be architected only with complete understanding of the attack surface of AMI. It has been demonstrated that a compromised AMI can severely disrupt the power grid and all critical infrastructure that rely on power. It is important to note that AMI is a global initiative with eventual deployment in any nation with a managed power grid. The first step in mitigating these attacks is to catalog the devices and protocols in order to scope the attack surface. This will establish the knowledge required to begin the process of reducing this attack surface through resiliency of the exposed points. In this paper, the attack surface has been established with the inclusion of some of the most prominent attack vectors into this surface.